\documentclass{jps-cp}
\usepackage{txfonts} 
\usepackage{wrapfig}
\usepackage{epsfig}

\title{Status of the Laser Spectroscopy and Merged-beam Experiments at RICE}

\author{Y. \textsc{Nakano}$^{1,2}$, R  \textsc{Igosawa}$^{3}$, S  \textsc{Iida}$^{1}$, S  \textsc{Okada}$^{2}$, M  \textsc{Lindley}$^{2}$, S \textsc{Menk}$^{1,2,4}$, R. \textsc{Nagaoka}$^{1}$, \\ T. \textsc{Hashimoto}$^{5}$, S. \textsc{Yamada}$^{4}$, T. \textsc{Yamaguchi}$^{3}$, S. \textsc{Kuma}$^{2}$, T  \textsc{Azuma}$^{2,4}$}

\inst{$^{1}$Department of Physics, Rikkyo University, Tokyo 171-8501, Japan \\
$^{2}$AMO Physics Laboratory,  RIKEN,  Saitama 351-0198, Japan\\
$^{3}$Department of Physics, Saitama University, Saitama 338-8570, Japan\\
$^{4}$Department of Physics, Tokyo Metropolitan University, Tokyo  192-0397, Japan\\
$^{5}$Japan Atomic Energy Agency (JAEA), Tokai 319-1184, Japan
}
\email{ nakano@rikkyo.ac.jp}

\recdate{February 28, 2018}

\abst{Recently, we reported the commissioning of the new cryogenic ion storage ring RICE, which demonstrated potential capabilities for the precise studies of molecular structures and reaction dynamics. 
In the present article, we describe the status of experimental programs ongoing at RICE with a focus on the laser spectroscopy and merged-beam collision experiments. }

\kword{Electrostatic storage ring, laser spectroscopy, neutral beam, TES~microcalorimeter}

\begin{document}
\maketitle

\section{Introduction}

The RIKEN Cryogenic Electrostatic ion storage ring (RICE) has been built, and its operation has been tested at cryogenic temperatures down to 4.2~K \cite{Nakano2017}. 
The commissioning experiments with a 15~keV-Ne$^+$ ion beam demonstrated a storage lifetime of $\sim$ 780~s, and the residual gas density was estimated at a few 10$^4$~cm$^{-3}$. 
In addition to providing such an extreme vacuum condition, the cryogenic cooling of the ring opens a new opportunity to study the molecular structure and reaction dynamics in the absence of excitation by blackbody radiations. 
Radiative cooling of the molecular ions into the rovibrational ground state during storage have already been demonstrated in the cryogenic storage rings DESIREE \cite{Schmidt2017} and CSR\cite{OConnor2016,Meyer2017}.

With the aid of the new capabilities provided by the cryogenic storage ring, several experimental programs are currently running at RICE. 
Among all the experiments that are going to be performed at RICE, probing the internal temperature of stored molecules is one of the most important tasks.   
Therefore, a dissociation spectroscopy of molecular ions is currently being performed for the direct and real-time observation of the rovibrational population of molecules during storage. 
In addition, to prepare for injections of pre-cooled molecular ion beams, a cryogenic ion trap \cite{Menk2018} and a helium droplet source \cite{Kuma2017} are being developed in parallel. 
The ion trap injection beamline is currently equipped with an electrospray ionization source for large molecular ions and a laser ablation source for clusters ions. 
Ions from either source were successfully accumulated in the cryogenic linear octupole trap, and then extracted and accelerated to $\sim$10~keV for injection. 
The ion trap beamline is already connected to the main beamline of the RICE via a quadrupole deflector, and ion injection is under commissioning.


\begin{figure}[tb]
\includegraphics[width=1\linewidth]{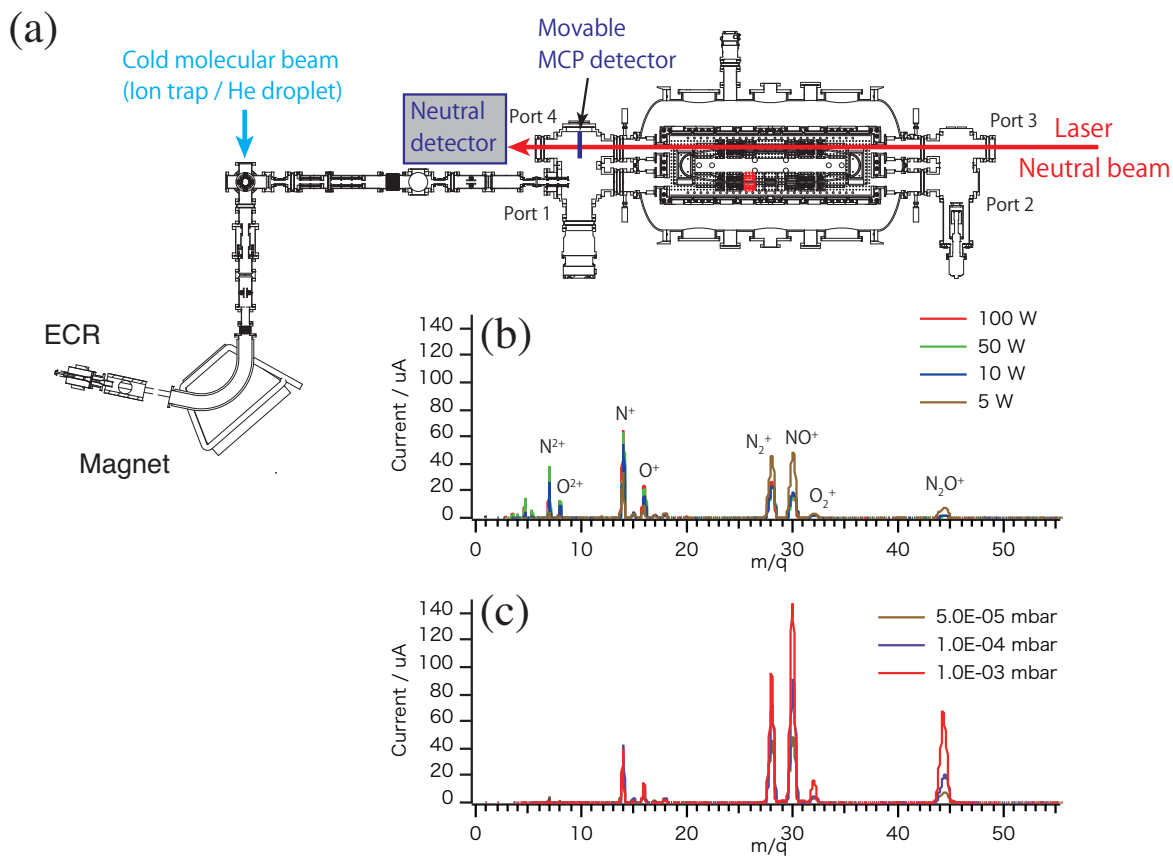}
\caption{(a) Schematic view of the RICE beam line configuration. Laser and/or neutral beams are injected from Port 3, and the neutral particle detector is installed at Port 4. (b) Mass spectrum obtained from the ECR ion source operated at different microwave powers, namely, 5, 10, 50, and 100~W. The source gas was N$_2$O and the source pressure was 5.0$\times 10^{-5}$~mbar. (c) The same but measured at different source pressures. The microwave power was kept at 5~W. }
\label{fig:beamline}
\end{figure}

Another important part of the facility is the laser and neutral beam sources. 
As shown in Fig.~\ref{fig:beamline}(a), the laser and neutral beams will be introduced from Port 3 to be merged with the stored ion beam at the straight section. 
Tunable laser systems, visible-UV OPO, IR OPO, and dye lasers are installed together with necessary control and detection systems for the planned laser experiments at RICE. 
The neutral beamline has already been constructed, and it is under off-line testing with negative ion sources and a photo-detachment laser at Rikkyo University. 
In addition, the development of an energy-sensitive neutral detector has been started. 
In the following, the current status of these experiments and developments will be reported.

\section{Laser dissociation experiments}
\subsection{Storage of molecular ions}
Molecular ions were produced by an ECR ion source described in Ref.\cite{Nakano2017}. 
The ECR ion source is designed for high current beams of multiply charged atomic ions, and is generally operated with a high microwave power for maintaining a stable plasma. 
However, running the source with a low microwave power enables molecular ions to remain in the source plasma without being broken up. 
Figure \ref{fig:beamline}(b) shows the mass spectrum from the ECR source operated with $5\times 10^{-5}$ mbar N$_2$O gas under different values of microwave power, namely 5, 10, 50, and 100~W. 
Under the high microwave input, the mass corresponding to N$_2$O$^+$ was not observed but its fragments dominated the mass population.
When the input microwave power was reduced to 10 and 5~W, the intensity of the fragment ions decreased while the N$_2$O$^+$ beam current increased. 
The source pressure dependence of the mass spectra under a low microwave power of 5~W is shown in \ref{fig:beamline}(c). 
The intensity of the N$_2$O$^+$ beam was enhanced by increasing the N$_2$O gas flow into the source. 
Under a source pressure of $1\times 10^{-3}$ mbar, the N$_2$O$^+$ beam current was around 70 $\mu$A. 
After passing through a $2\times 2$~mm slit and following beam transport sections, the beam current at the injection was 10~nA.


\subsection{Laser setup}
A narrow-band pulsed dye laser was installed at Port 3 of the storage ring (see Fig.~1). 
For the dissociation spectroscopy of N$_2$O$^+$ ions at a laser wavelength of around 320~nm, the second harmonics from the 532~nm pumped DCM/Ethanol dye laser output is used. 
The pumping source was recently upgraded from a flashlamp-pumped 10~Hz Nd:YAG laser to a high-repetition DPSS Nd:YAG laser with a pulse energy of 80~mJ at 100~Hz. 
The typical pulse energy at the output of the second harmonics generator is 5 mJ. 

\begin{figure}[tb]
\includegraphics[width=1\linewidth]{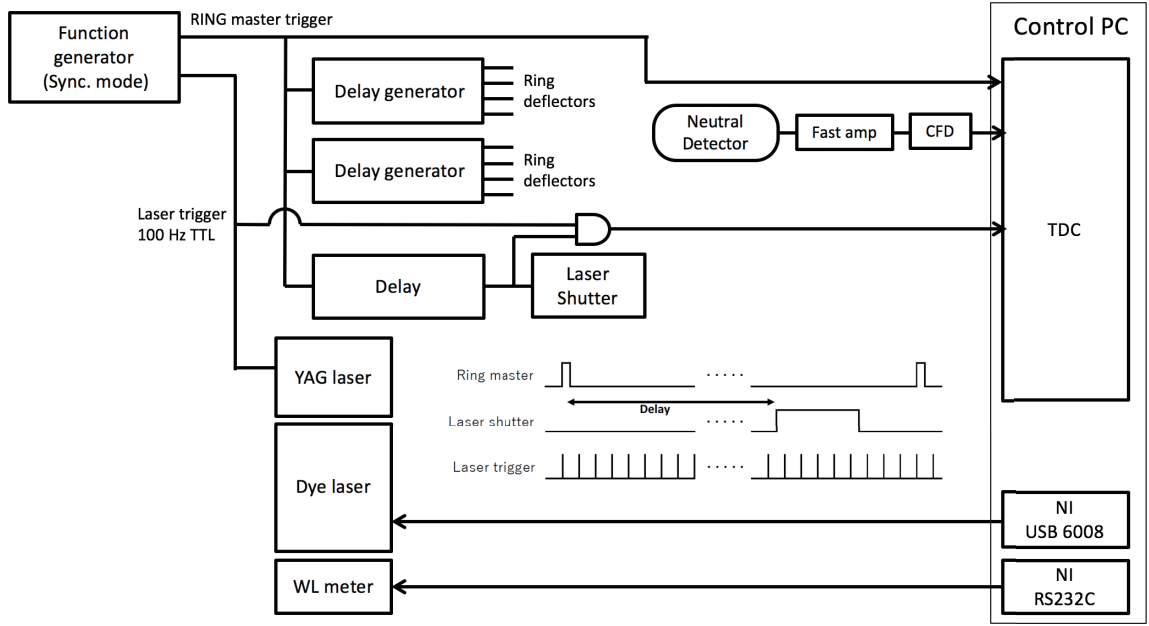}
\caption{Control and timing chart of the laser and data taking system. }
\label{fig:timing}
\end{figure}

Fig.~\ref{fig:timing} shows the control and timing sequence of the storage ring and laser system. 
The master trigger for the ring injection cycle and external Q-switch trigger for the laser were generated by a two-channel function generator in a synchronized mode. 
While the laser is constantly triggered at 100~Hz, the repetition cycle of the ring master trigger depends on the experiments, \textit{e.g.}, more than thousands seconds for long-period storage measurements. 
The laser light is introduced into the ring through a mechanical shutter, which opens at a certain delay time after the ion injection into the ring. 
The logical conjugation of the laser trigger and shutter signal was sent to the DAQ PC as the signal of the laser irradiation into the ring.

\subsection{Neutral particle detector}

\begin{figure}[tbh]
\begin{center}
\includegraphics[width=1\linewidth]{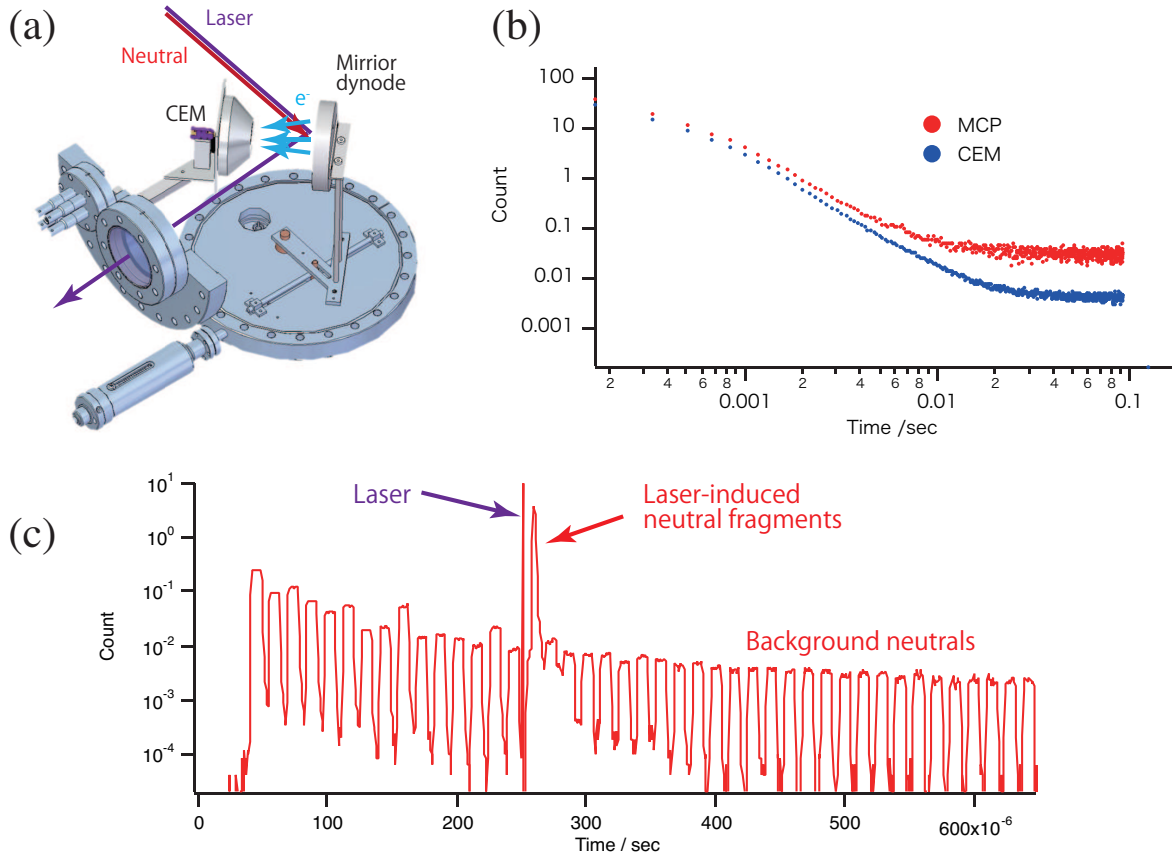}
\caption{(a) A three-dimensional view of the neutral particle detector. (b) Decay curve of the 15~keV~N$_2$O$^+$ ion beam in the RICE observed by the movable MCP detector (red) and the CEM of the mirror dynode detector (blue).  The vertical axis represents the normalized count per TDC channel with a 167~$\mu$s bin size. (c) The same measurement by the mirror dynode detector but shown in a different bin size of 655~ns. A pulsed laser was introduced at 250~${\mathrm \mu}$s after the injection.}
\label{fig:detector}
\end{center}
\end{figure}

In the laser dissociation experiment, the laser beam is merged with the stored ion beam at the straight section of the ring. 
Laser-induced dissociation of molecular ions such as
\begin{equation}
AB^+ + h\nu \rightarrow A^+ + B
\end{equation}
can be identified by detecting the neutral fragment that flies straight to the outside of the ring through Port 4. 
Microchannel plates (MCP) are widely used as detectors for keV neutral particles in ion beam experiments. 
However, the direct UV laser irradiation will immediately lead to an irreversible damage on the MCP surface. 
In this respect, MCPs are not suitable for the collinear laser experiments, and therefore, a dedicated detection system has been developed. 
The detection method is based on the conversion of a neutral particle into secondary electrons on a dynode surface. 
As shown in Fig.~\ref{fig:detector}(a), the laser and neutral particles impinge at 45 degrees on the conversion dynode, which is a pure aluminum plate finely polished to a mirror surface. 
Because aluminum has a high reflectance in the UV region, more than 90~\% of the laser light is reflected to the outside of the chamber through the fused silica window. 
On the other hand, hits of the fast neutral particles produce secondary electrons from the dynode. 
These electrons are then accelerated towards the channel electron multiplier (CEM) installed in front of the dynode. 

The mirrored dynode was installed at Port 4 of the RICE, $\sim$450~mm behind the movable MCP detector, and its efficiency was evaluated without lasers. 
Figure~\ref{fig:detector}(b) shows the decay curves of the 15-keV N$_2$O$^+$ ion beam measured by the MCP and CEM of the mirror dynode detectors.
Because the dark count rate of CEM is lower than that of the MCP, the CEM measurement showed a decay curve with a better signal-to-noise ratio. 
Considering the fact that the opening ratio of the MCP is around 60~\%, the mirror dynode detector is expected to have a higher detection efficiency than the MCP. 
The collection efficiency of the secondary electrons to the CEM is presently under the process of improvement. 


The response of the mirror dynode detector to the laser shot was tested with the N$_2$O$^+$ ion beam the and the 323~nm dye laser pulse. 
The time spectrum of the CEM signal is shown in Fig.~\ref{fig:detector}(c), where the time origin corresponds to the beam injection.  
Each bunch of the neutral signal corresponds to the circulation of the ion bunch in the ring. 
The laser was fired 250~${\mathrm \mu}$s after the injection, which created a sharp peak on the spectrum by the scattered laser light and photoelectrons from the dynode. 
Around 10~${\mathrm \mu}$s after the laser was fired, the neutral yield was enhanced by two orders of magnitude due to the laser-induced dissociation of the N$_2$O$^+$ ions. 
This delay corresponds fo the time of flight of neutral particles from the reaction region to the detector. 
Counting the number of laser-induced signals as a function of the laser wavelength enables the in-ring dissociation spectroscopy of molecular ions. 




\section{Neutral beam injection}
\subsection{Negative ion source}
\begin{figure}[tb]
\includegraphics[width=1\linewidth]{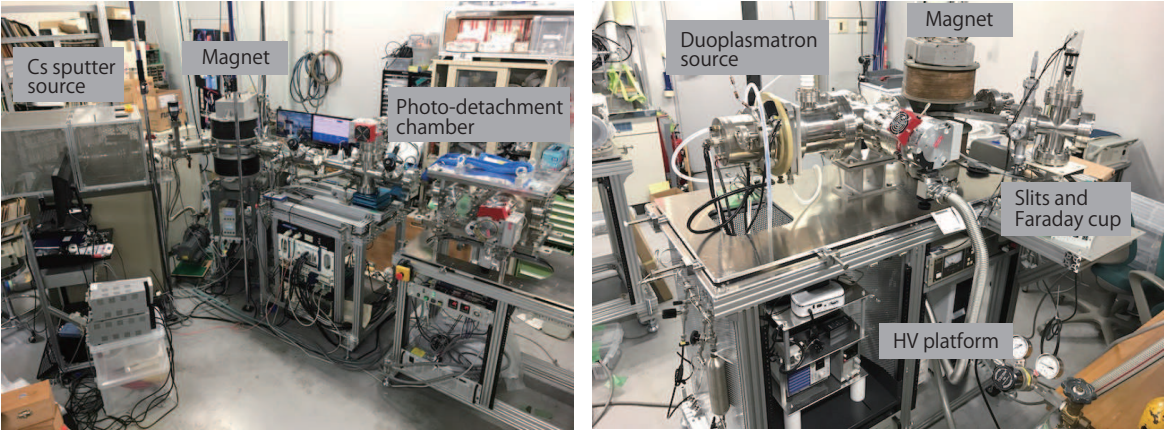}
\caption{Photographs of the test beam line for the Cs sputter (left), and the duoplasmatron (right) ion sources. }
\label{fig:source}
\end{figure}

The neutral beam is generated by a laser-induced electron detachment of a negative ion beam \cite{OConnor2015a}. 
Two different ion sources, the duoplasmatron and the cesium sputter ion sources, were installed on the beamline at Rikkyo University for the off-line testing (Fig.~\ref{fig:source}). 
The cesium sputter ion source (SNICS II, National Electronics Corp.) was prepared for solid targets, such as carbon, silicon, and sulfur, and tested with $^{12}$C at different extraction energies ranging from 5 to 20 kV. 
By using a set of beam profile monitors (BPM80, National Electronics Corp.), the source operation and beam transport parameters were optimized. 
An additional beam steerer was installed after the dipole magnet to compensate for the misalignment of the source and the magnet. 
At present, several microamp of C$^-$ beam is transported to the end of the test beamline. 

For gaseous targets like hydrogen, deutrium, and oxygen, a duoplasmatron ion source (High Voltage Engineering Model 358) was also set up as an injector. 
The source has been installed on a 30~kV high-voltage platform with the driving power supplies controllable via a wireless network, and it is connected to a 100-mT dipole magnet followed by an XY slit and a Faraday cup for mass analysis. 
Following the computer simulations of the beam trajectory, the source operation is currently under testing using noble gases.

\subsection{Photo-detachment laser}

An 808~nm high-power diode laser array is used to neutralize the negative ion beam by photo-detachment. 
The laser head is a stack of 60 diode bars with a continuous output power of 5~kW in total. 
A 15~kW DC power supply and deionized water circulating system with a cooling power of 7~kW were installed to drive the laser. 
The power supply is securely interlocked by water flow and temperature monitors. 
The output power of the laser was measured using a plano-convex lens and a high-power laser probe (FLASH-6K, GENTEC) as a function of the electric current of the power supply. 
The laser threshold current was $\sim$18~A, and the operating current at 5~kW was $\sim$92~A. 

\begin{figure}[tb]
\includegraphics[width=1\linewidth]{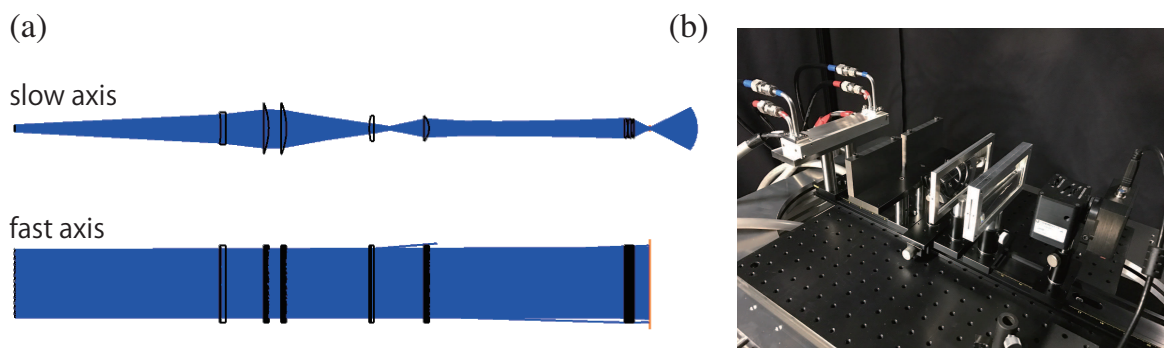}
\caption{(a) Example of the laser ray tracing simulation using Zemax. The upper and lower figures show the results for the fast and slow axes, respectively. (b) A photograph of the laser test bench. }
\label{fig:lda}
\end{figure}

The diode array has a large emission area of 100 mm $\times$ 12 mm and a large angular divergence of $\sim$40 and $\sim$5 degrees in the fast and slow axes, respectively. 
With the use of a microlens array, the fast axis divergence was reduced to less than  0.5~degree.  
However, a well-designed optical system is still needed to increase the efficiency of photo-detachment.
We carried out a laser ray-tracing simulation using Zemax Optic Studio. 
Figure \ref{fig:lda}(a) shows an example of the simulated laser ray in the fast and slow axes. 
The beam width is converged to 1.2 mm by using three convex cylindrical lenses. 
The real optics are constructed on a laser test bench (Fig.~\ref{fig:lda}(b)) with multiple ND filters and a laser beam profiler. 
A multi-path amplification of the laser light in a cavity is under preparation.

\section{Development of energy-sensitive neutral particle detector}
\subsection{TES microcalorimeters for mass spectrometric identification of neutral molecules}

Identification of molecular products in the merged-beam experiment is important for achieving a comprehensive understanding of the reaction mechanism. However, neither the MCP nor the mirror dynode detector described above has sufficient energy resolution to measure the energy deposit for the mass spectrometric identification.

The Transition-Edge-Sensor (TES) microcalorimeter is a cryogenic detector that takes advantage of superconductivity \cite{Ens05}. TES measures the tiny temperature rise resulting from the interaction with an incident particle due the abrupt increase in the resistance of a thin superconducting film that is biased within a sharp transition edge between the normal and superconducting phases. This results in a high energy resolution. Though being developed for photon detection, e.g. for X-rays \cite{Randy17}, TES has sufficient sensitivity for any particle whose energy can be converted into heat. Here, we aim to apply this detector to the low-energy ($\sim$ 10 keV) molecule measurement at RICE.

\begin{figure}[tb]
\begin{center}
\includegraphics[width=0.5\linewidth]{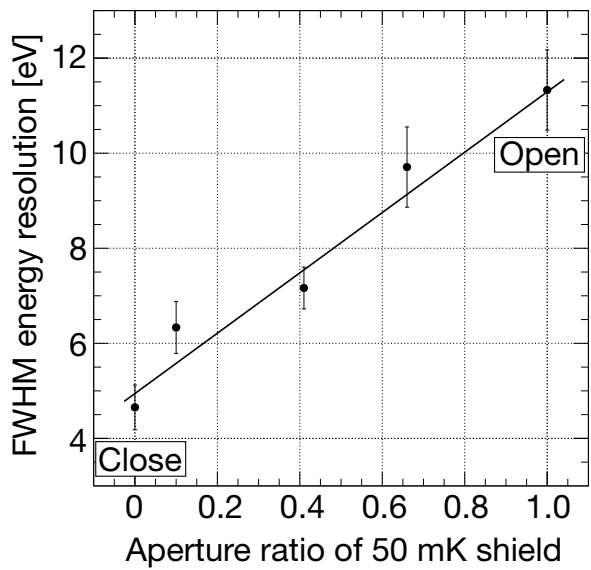}
\caption{FWHM energy resolution of the TES spectrometer for 6-keV X-ray as a function of the aperture ratio of the 50 mK shield. The error bar shows the RMS of the dispersion for each TES pixel.}
  \label{Fig:resolution}
  \end{center}
\end{figure}

%
%
\subsection{Energy resolution of the TES microcalorimeter}

The TES is normally operated at 100 mK being shielded from heat radiation using infrared blocking filters made of thin aluminum foils at each of the three cooling stages, 50 mK, 3 K, and 50 K \cite{HeatesPSI}. Unlike X-rays, the low-energy molecules cannot pass through the filters made of 100-nm-thick aluminum foil. Here, it should be noted that the two shields at 3 K and 50 K are fortunately avoidable in RICE because it is cooled already at 4 K.

For this application, the effect of the 50 mK shield against the radiation background was evaluated from the Mn $K_\alpha$ ($\sim$ 6 keV) spectrum obtained with a $^{55}$Fe radioactive source for different configurations of the shield. The detector system is based on a 240-pixel TES array developed by NIST (the same system as that in Ref. \cite{HeatesPSI}), where only a part of the pixels is used. Under normal conditions, with a 50 mK shield having a 5-$\mu$m-thick aluminum foil, the FWHM energy resolution of a summed spectrum of 23 working TES pixels was 4.7 eV. When the shield is completely removed (being fully open to the 3K radiation), the TES system still functions with an energy resolution of 11.3 eV. Figure \ref{Fig:resolution} shows those results together with results of the measurements with copper foils having a mesh structure with different aperture ratios. In summary, the condition can be chosen to optimize a trade-off between energy resolution and aperture ratio (efficiency for the incident particle). Following this study, we are preparing an actual setup for molecular-beam measurement in RICE towards the practical use of the merged-beam experiment.

\section{ Acknowledgement}
This work was supported by the RIKEN Basic Science Interdisciplinary Research Projects, and by the JSPS KAKENHI Grant Number 26220607.
Y.N. acknowledges the financial support from JSPS KAKENHI (No. 16K13859 and 17H02993). 
S.K. and S.O. thank the members of the HEATES collaboration for their support and discussions towards this challenging application of TES to the low-energy molecule measurement, and acknowledge the financial support from Yamada Science Foundation and JSPS KAKENHI (Nos. 26707014, 16H02190).
S.Y. acknowledges the financial support from JSPS KAKENHI (No.15H05438). 
T.H acknowledges the financial support from JSPS KAKENHI (No.16K17718). 
This work was partially supported by MEXT-Supported Program for the Strategic Research Foundation at Private Universities, 2014-2017 (S1411024).


\begin{thebibliography}{9}

\bibitem{Nakano2017} Y. Nakano, Y. Enomoto, T. Masunaga, S. Menk, P. Bertier, and T. Azuma, Rev. Sci. Instrum. {\bf 88}, 33110 (2017).

\bibitem{Schmidt2017} H. T. Schmidt, G. Eklund, K. C. Chartkunchand, E. K. Anderson, M. Kami\'{n}ska, N. De Ruette, R. D. Thomas, M. K. Kristiansson, M. Gatchell, P. Reinhed, S. Ros\'{e}n, A. Simonsson, A. K\"{o}llberg, P. L\"{o}fgren, S. Mannervik, H. Zettergren, and H. Cederquist, Phys. Rev. Lett. {\bf 119}, 073001 (2017).

\bibitem{OConnor2016} A. P. O'Connor, A. Becker, K. Blaum, C. Breitenfeldt, S. George, J. G\"{o}ck, M. Grieser, F. Grussie, E. A. Guerin, R. von Hahn, U. Hechtfischer, P. Herwig, J. Karthein, C. Krantz, H. Kreckel, S. Lohmann, C. Meyer, P. M. Mishra, O. Novotn\'{y}, R. Repnow, S. Saurabh, D. Schwalm, K. Spruck, S. Sunil Kumar, S. Vogel, and A. Wolf, Phys. Rev. Lett. {\bf 116}, 113002 (2016).

\bibitem{Meyer2017} C. Meyer, A. Becker, K. Blaum, C. Breitenfeldt, S. George, J. G\"{o}ck, M. Grieser, F. Grussie, E. A. Guerin, R. Von Hahn, P. Herwig, C. Krantz, H. Kreckel, J. Lion, S. Lohmann, P. M. Mishra, O. Novotn\'{y}, A. P. O'Connor, R. Repnow, S. Saurabh, D. Schwalm, L. Schweikhard, K. Spruck, S. Sunil Kumar, S. Vogel, and A. Wolf, Phys. Rev. Lett. {\bf 119}, 023202 (2017).


\bibitem{Menk2018}S. Menk, P. Bertier, Y. Enomoto, T. Masunaga, T. Majima, Y. Nakano, and T. Azuma, Rev. Sci. Instrum. {\bf 89}, 113110 (2018).



 \bibitem{Kuma2017} S. Kuma and T. Azuma, Cryogenics. {\bf 88}, 78 (2017).



\bibitem{OConnor2015a} A. P. O'Connor, F. Grussie, H. Bruhns, N. de Ruette, T. P. Koenning, K. A. Miller, D. W. Savin, J. St\"{u}tzel, X. Urbain, and H. Kreckel, Rev. Sci. Instrum. {\bf 86}, 113306 (2015).

 \bibitem{Ens05} 
	 K.D. Irwin and G.C. Hilton, ``Transition-Edge Sensors'', 	
	 C. Enss (ed.), Cryogenic Particle Detection,
         Topics in Applied Physics, vol. {\bf 99}, Springer, 2005.

 \bibitem{Randy17} 
	 W.D. Doriese et al., Rev. Sci. Instrum. {\bf 88} (2017) 053108. 

 \bibitem{HeatesPSI} S. Okada et al., Prog. Theor. Exp. Phys. {\bf 2016}, 091D01.
 

 
\end{thebibliography}
\end{document}